\documentclass[10pt,a4paper]{article} 
\usepackage[dvips]{graphicx}
\usepackage{amsfonts,amsmath,amssymb}
\usepackage{hyperref}
\usepackage{cite}
\newcommand{\bse}{\begin{subequations}}
\newcommand{\ese}{\end{subequations}}

\begin{document}
\title{\bf Magnetized  Einstein-Maxwell-dilaton model under external electric fields}

\date{}
\maketitle
\vspace*{-0.3cm}
\begin{center}
{\bf Leila Shahkarami}\\
\vspace*{0.3cm}

{\it {School of Physics, Damghan University, Damghan, 41167-36716, Iran
}} \\
\vspace*{0.3cm}
{\it  {l.shahkarami@du.ac.ir}}
\end{center}

\begin{abstract}
We employ an analytic solution of a magnetized Einstein-Maxwell-dilaton gravity system whose parameters have been determined so that its holographic dual has the most similarity to a confining QCD-like theory influenced by a background magnetic field. Analyzing the total potential of a quark-antiquark pair in an external electric field, we are able to investigate the effect of the electric field on the different phases of the background which are the thermal AdS and the black hole phases. 
This is  helpful for better understanding the confining character and also the phase transitions of the system. 
We find out that the field theory dual to the black hole solution is always deconfined. 
However, although the thermal AdS phase describes the confining phase in general, for the quark pairs parallel to $B$ (longitudinal case) and $B>B_{\mathrm{critical}}$ the response of the system to the electric field mimics the deconfinement. 
We moreover consider the effect of the magnetic field and the chemical potential on the Schwinger effect. 
We observe that when we are in the black hole phase with sufficiently small values of $\mu$ or in the thermal AdS phase, and for both longitudinal and transverse cases, the magnetic field increase leads to the enhancement of the Schwinger effect, which can be termed as the inverse magnetic catalysis. 
This is deduced both from the decrease of the critical electric fields and from the decreasing the height and width of the total potential barrier that the quarks are facing with. 
However, by increasing $\mu$ to high enough values, the inverse magnetic catalysis turns into magnetic catalysis, as can also be observed from the diagram of the Hawking-Page phase transition temperature versus $B$ for the background geometry itself.
\end{abstract}
\section{Introduction}
The Schwinger effect \cite{Schwinger} is a fascinating nonperturbative phenomenon expected in Quantum Electrodynamics (QED), where electron-positron pairs are created out of the vacuum due to the presence of an external electric field. 
This effect is not restricted to QED, but is a general feature of any quantum field theory equipped with a $U(1)$ gauge field, e.g., quarks and antiquarks have electric charges and hence can be produced as pairs from the QCD vacuum when the electric force acting on the quarks overcomes the confining force between them.

The first calculation in the context of the Schwinger effect was done in the small field strength and small coupling approximation in \cite{Schwinger} resulted in an exponentially suppressed expression for the production rate of particles, expressing the phenomenon as a quantum tunneling process.
This was later generalized to the arbitrary coupling case \cite{manton1}, where one can deduce a critical electric field strength, $E_c$, below which the Schwinger effect occurs only through a tunneling process, but as the electric field strength exceeds $E_c$, the pair of charged particles are produced freely without any obstacle. 
However, the existence of such a critical behavior was not clarified by these calculations, since the value found for $E_c$ does not satisfy the weak-field condition. 

Motivating by the existence of a similar critical behavior in the presence of the electric field in the string theory \cite{string1,string2}, people hired the AdS/CFT correspondence \cite{Maldacena2,Maldacena3,Maldacena1,Solana} and confirmed the critical behavior by a series of studies referred to as the holographic Schwinger effect. 
Semenoff and Zarembo \cite{semenoff} for the first time calculated the Schwinger pair production by the use of AdS/CFT and found the critical electric field.
Following this work, the holographic Schwinger effect has drawn a lot of interests where many different aspects of the problem were put under study, by employing various theories with different properties such as the existence of the confinement, magnetic field, chemical potential, etc. 
Studies in the confining phase reveals the existence of a second critical electric field, $E_s$, which is the minimum value necessary for the Schwinger effect to occur.
There are two main approaches for most of these investigations.
One of them \cite{semenoff,potential,Sch1,Sch2,Sch3,confin1,confin2,Sch5,Sch6,Sch7,
confinrev,dehghani,davoud,magnetized} contains the interpretation of the Schwinger effect using the DBI-action of probe branes in AdS/CFT, where they analyze the total potential of a particle-antiparticle pair in the presence of electric fields using the rectangular Wilson loop and/or calculate the production rate by the use of the temporal circular Wilson loop.
The other approach \cite{decay1,decay2,equen,magneticdecay,LF} concentrates on the flavor branes in holographic QCD, where they investigate the instability imposed by the electric fields by calculating the effective Lagrangian of the theory under an electric field, holographically equivalent to the imaginary part of the effective DBI-action of the flavor probe brane.

The importance of the study of the nonperturbative Schwinger effect is twofolded.
 Although it has not been observed in experiments since the minimum electric field necessary for the effect to be detectable is out of reach of the strongest lasers available, the Schwinger effect is a highly-accepted realistic prediction of the quantum field theories, promised to be accessible by the high-intensity lasers in near future \cite{exper1,exper2}.
Hence, it deserves to be studied both theoretically and experimentally, as a real phenomenon as it plays an important role, e.g, in heavy ion collisions and magnetors (dense neutron stars).
On the other hand, the study of this effect can be very informative for theoreticians because of its nonperturbative nature which makes it useful for understanding the vacuum structure and the critical behavior specially in confining theories.  

Although the holographic response of a confining field theory has been extensively put under study using various holographic theories, the studies with models that mimic the desired QCD features are rare.
Top-down models having the important roles in these investigations are not quite appropriate in describing the real QCD.
The phenomenologycal bottom-up AdS/QCD models are better candidates since they can be tuned to match many considerably experimental and lattice QCD results, although they do not enjoy the validity of the duality to the extent of the top-down models.
There are an advanced category of the bottom-up models with a better match to real QCD, which are explicit solutions of the equations of motion, unlike the other ones and some of which also enjoy a running coupling constant.
As far as we know, the use of such models is very rare and not enough to capture all the aspects of the QCD-like theories in response to extreme external electric fields (as examples of the deconfined phase, see \cite{GB,magnetized}).

In many important physical situations a magnetic field is present along with an electric field, an example of which is seen in the RHIC and LHC experiments where strong electromagnetic fields are created at the very stage of the heavy-ion collisions \cite{sem1,sem2,sem3,sem4} and in spite of the rapid decay of the magnetic field after the collision, it perhaps remains sufficiently large till the quark-gluon plasma (QGP) forms \cite{Blife1,Blife2}.
Therefore, the study of the behavior of a QCD-like theory under the simultaneous influence of electric and magnetic fields can be very fruitful.
Some examples of such investigations can be found in the literature \cite{Sch2,davoud,magnetized,decay2,magneticdecay}, which consider different aspects of the problem in different holographic setups.
In almost all of these studies, the back reaction of the magnetic field on the background geometry is neglected. 
As far as we know only in \cite{magnetized} the response of a magnetized field theory in the deconfined phase to an external electric field has been studied. 
Our aim in this paper is to investigate the effect of an external constant electric field on a magnetic field dependent confining gauge theory, dual to the gravity solution obtained in \cite{hajilou}. 
They employed an Einstein-Maxwell-dilaton (EMD) gravity system with two Maxwell fields and obtained an exact solution of this system via the potential reconstruction method (see e.g., \cite{hajilou,reconst1,reconst2,reconst3,reconst4,reconst5} and references therein), incorporating a magnetic field and a running dilaton.
This complete solution, also equipped with the finite temperature and the chemical potential, is expressed in terms of a single scale function $A(z)$ chosen at will and some free parameters determined by comparing the holographic QCD results to the lattice QCD ones with vanishing magnetic field and chemical potential.
Following them we also choose the simplest form for the scale function, i.e., the quadratic dependence to the holographic radial coordinate.
With the aid of this setup, we are able to investigate the Schwinger effect in a confining QCD-like theory which has the most similaty to the real QCD and influenced by a background magnetic field.
We can also explore the effect of the anisotropy induced by the magnetic field on the Schwinger effect in different directions.
Another motivation is that the study of the critical behavior of the confninig system when an external electric field is applied could be helpful in revealing the (de)confining property of the theory.

Next section is devoted to a short introduction of the model and its solution, where we also briefly review the phase structure of the gravity side by studying the thermodynamics of the system.
Then, we apply an external constant electric field on this model in Sec.\,3 and study the response of the system by analyzing the quark-antiquark potential.
Finally, concluding remarks and discussion are presented in the last section.


\section{Einstein-Maxwell-dilaton system with a magnetic field}
In this section we present an introductory review of the background geometry of our interest, constructed in \cite{hajilou}.

\subsection{Background geometry}
The action of the 5-dimensional EMD gravity system containing two Maxwell fields in the Einstein frame is given by \cite{hajilou}
\begin{align}\label{EMD}
  S_{\mathrm{EMD}}=-\frac{1}{16 \pi G_5}\int d^5 x \sqrt{-g} \left[R -\frac{f_1(\phi)}{4}F_{(1)}^2-\frac{f_2(\phi)}{4}F_{(2)}^2-\frac{1}{2}\partial_{\mu}\phi \partial^{\mu}\phi-V(\phi)\right],
\end{align}
where $\phi$ is the dilaton field with the potential $V(\phi)$.
Also, $F_{(1)\mu \nu}$ and $F_{(2)\mu \nu}$ are the field strength tensors, and $f_1(\phi)$ and $f_2(\phi)$ are the kinetic gauge functions representing the coupling between the gauge fields and the dilaton field.
These two gauge fields are employed to introduce the chemical potential and a constant magnetic field to the dual theory.

Now we present the complete magnetized black brane solution with running dilaton for the above system, obtained in \cite{hajilou}.
The metric in the string frame satisfying the equations of motion with the desired boundary conditions reads
\begin{align}\label{metric}
  ds_s^2=\frac{L^2 e^{2{\cal A}_s}}{z^2}\left[-g(z)dt^2+\frac{dz^2}{g(z)}+dx_{\parallel}^2+e^{B^2z^2}\left(dx_{\perp 1}^2+dx_{\perp 2}^2\right)\right],
\end{align}
where $L$ is the AdS radius, $B$ is the background magnetic field in the $x_{\parallel}$ direction and $z$ is the holographic radial direction running from $z=0$ at the boundary to $z=z_h$ at the black hole horizon.
Moreover, choosing the warp factor of the metric in the Einstein frame as $e^{2A(z)}$, ${\cal A}_s(z)=A(z)+\frac{1}{\sqrt{6}} \phi(z)$.
This relation is found by using the dilaton transformation, i.e., $g_{s\mu \nu}=e^{\sqrt{\frac{2}{3}}\phi}g_{\mu \nu}$ in which $g$ and $g_s$ denote the metric in the Einstein frame and string frame, respectively.
The blackening function $g(z)$ is as follows:
\begin{align}\label{black}
g(z)=1+\int_0^z du\ u^3e^{-B^2u^2 -3A(u)}\left(K+\frac{\tilde{\mu}^2}{2cL^2}e^{cu^2} \right),
\end{align}
where $\tilde{\mu}=\frac{2c\mu}{e^{cz_h^2}-1}$ in which $\mu$ denotes the chemical potential in the field theory side and
\begin{align}\label{k}
K=-\frac{1+\frac{\tilde{\mu}^2}{2cL^2}\int_0^{z_h}du\ u^3 e^{-B^2 u^2 -3A(u)+cu^2}}{\int_0^{z_h}du\ u^3 e^{-B^2 u^2 -3A(u)}},
\end{align}
and the dilaton field is found as
\begin{align}\label{phi}
\phi(z)=\int dz \sqrt{-\frac{2}{z}\left(3 z A''(z)-3 z A'(z)^2+6A'(z)+2B^4 z^3+2B^2 z \right)}-\phi_0,
\end{align}
where $\phi_0$ is determined by demanding $\phi(z=0)=0$.
The interested reader is referred to \cite{hajilou} to see the explicit forms of the other functions appearing in the solution of the magnetized EMD system.

The above equations express an infinite family of black hole solutions of the Einstein-Maxwell-dilaton gravity system, Eq.\,(\ref{EMD}), for different forms of the scale factor $A(z)$. 
We can have a complete self-consistent solution once the function $A(z)$ is fixed.
It is worth mentioning that the physical behavior of the dual field theory is determined by the exact form of the scale function $A(z)$ and even a slight deformation of the form factor can lead to considerable alterations in the physical features of the boundary gauge theory. 
An example is found in \cite{hajilou2} where they show how the confining behavior of the field theory alters by a small deformation of $A(z)$.

Here, following \cite{hajilou} we choose a simple form for the scale factor as $A(z)=-az^2$.
Therefore, the scalar field takes the following analytic form:
\begin{align}\label{phiAcz2}
\phi(z)=&\frac{9a-B^2}{\sqrt{6a^2-B^4}}\log \left(\frac{ \sqrt{6a^2z^2+9a-B^4z^2-B^2}+z\sqrt{6a^2-B^4}}{\sqrt{9a-B^2}}\right)\nonumber\\+&z\sqrt{6a^2z^2+9a-B^2\left(B^2z^2+1\right)}.
\end{align}
For our purposes there is no need to write the explicit forms of the other functions.
The dilaton field should be real-valued in the whole $z$-interval and from Eq.\,(\ref{phiAcz2}) one simply learns that this condition is satisfied only for $B^4\leqslant B_c^4= 6a^2$.
Fortunately, the same condition guarantees that the potential of the dilaton is bounded from above. 
There are no other restrictions for choosing the values of the parameters.

The solution presented here has two free parameters $c$ and $a$ which are determined by matching to physical and lattice QCD results at $B=0$, as $c=1.16\ GeV^2$ and $a=0.15\ GeV^2$. 
Therefore, $B_c\simeq 0.61\ GeV$.

\subsection{Phase structure of the background}
Here, we present a short survey of the thermodynamics of the introduced background and argue the possible phase transitions as the parameters change.

The black hole temperature and entropy of the system can be simply obtained as
\begin{align}\label{temp}
T=-\frac{z_h^3 e^{-3A(z_h)-B^2 z_h^2}}{4 \pi}\left(K+\frac{\tilde{\mu}^2}{2cL^2}e^{cz_h^2} \right),
\end{align}
and
\begin{align}\label{entropy}
S=\frac{e^{B^2 z_h^2+3 A(z_h)}}{4z_h^3}.
\end{align} 
Now, using the first law of thermodynamics in grand canonical ensemble, the free energy can be computed from $dF=-SdT$ as follows:
\begin{align}\label{freeE}
F=\int_{z_h}^{\infty}S(z_h)T'(z_h)dz_h,
\end{align} 
where the prime sign refers to the derivative with respect to $z_h$.

\begin{figure}[h]
\begin{center}
\includegraphics[width=6.8cm]{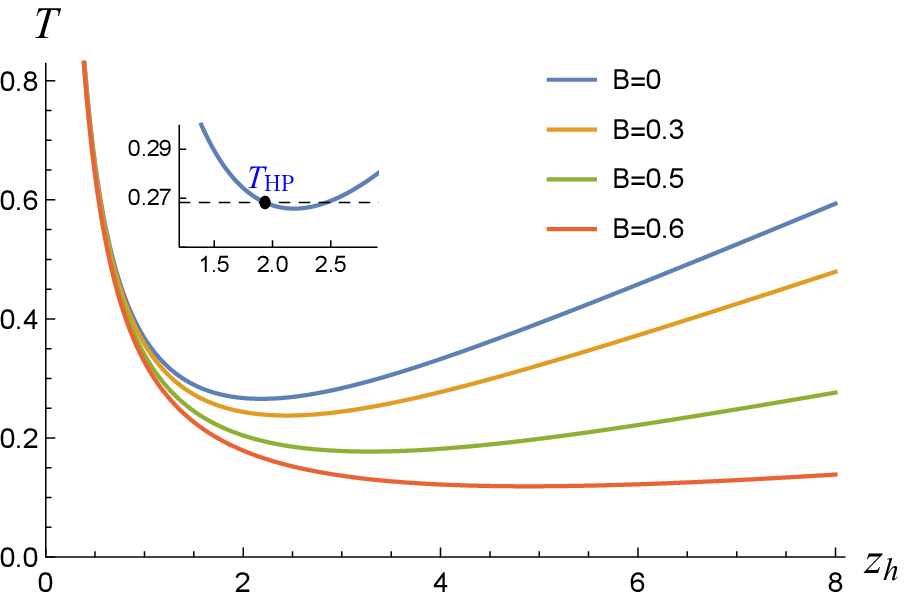}\hspace{0.3cm}
\includegraphics[width=6.8cm]{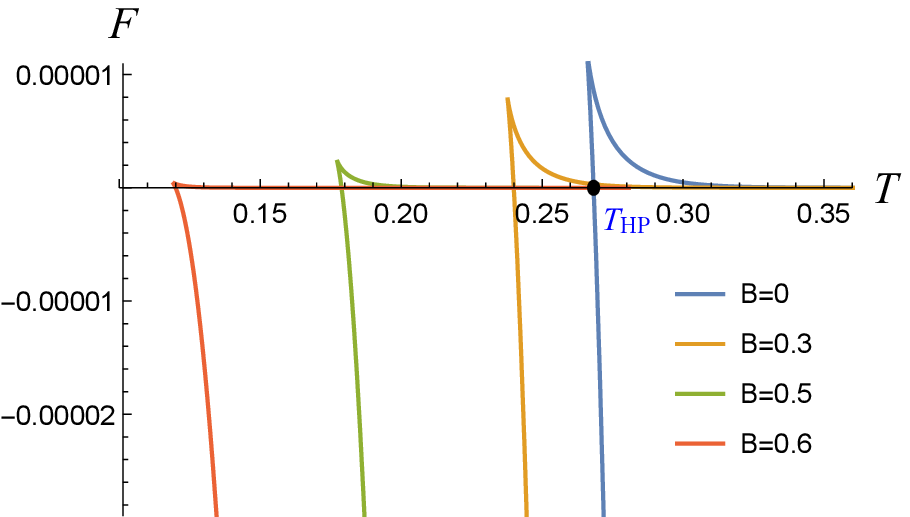}
\end{center}
\caption{\footnotesize 
Left and right graphs show the Hawking temperature as a function of the horizon position and the free energy as a function of the temperature, respectively, for various values of the magnetic field and $\mu=0$.}
\label{thermo}
\end{figure} 

\begin{figure}[h]
\begin{center}
\includegraphics[width=6.8cm]{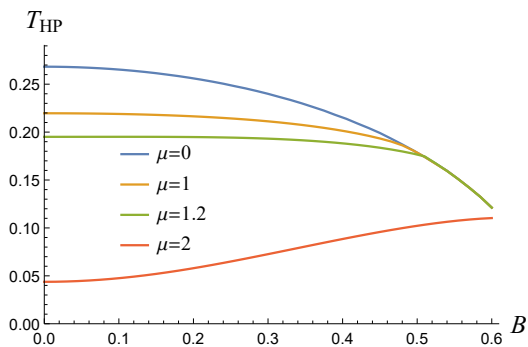}
\end{center}
\caption{\footnotesize 
The temperature of the Hawking-Page phase transition versus $B$ for various values of $\mu$.}
\label{T-HPB}
\end{figure} 

To see the behavior of these quantities at vanishing chemical potential, the graphs of the temperature versus the horizon radius and the free energy versus the temperature have been drawn in the left and right panels of Fig.\,\ref{thermo}, respectively.
Notice that in all the graphs, the values are in units $GeV$ which are not mentioned for the brevity of the notation.
In all the cases the temperature has a global minimum, $T_{\mathrm{min}}$, at some horizon radius, $z_{h,\mathrm{min}}$. 
For temperatures below $T_{\mathrm{min}}$ there exists no black hole solution.
However, for any value of $T>T_{\mathrm{min}}$, we are faced with two black hole solutions.
The black hole solution with $z_h>z_{h,\mathrm{min}}$ is unstable, for which the temperature increases as the black hole shrinks.
To reveal the stable/unstable nature of the solutions, one should study the free energy of the background as a function of the temperature (the right panel in Fig.\,\ref{thermo}).
As can be seen, the free energy graph intersects the horizontal axis at some value of the temperature, $T_{\mathrm{HP}}$, which is higher than $T_{\mathrm{min}}$ as shown in the zoomed box of the left panel in Fig.\,\ref{thermo} for $B=0$ as an example.
In fact, for a given set of parameters of the system, i.e., $\mu$ and $B$, only the large black hole solutions relating to $z_h<z_{h,\mathrm{HP}}$ are stable, for which the free energy is less than the small black hole solutions. 
Therefore, a Hawking-Page phase transition happens at $T_{\mathrm{HP}}$, i.e., the stable solution for $T>T_{\mathrm{HP}}$ is the large black hole solution with the horizon $z_h<z_{h,\mathrm{HP}}$ and negative free energy and the stable solution for $T<T_{\mathrm{HP}}$ is the thermal AdS solution with $z_h \to \infty$ and zero free energy.

Figure \ref{T-HPB} shows the behavior of the critical temperature $T_{\mathrm{HP}}$ as the magnetic field enhances, for some values of $\mu$. While the decrease of $T_{\mathrm{HP}}$ with $B$ is obviously observed for sufficiently small values of the chemical potential, this behavior is reversed for higher values of $\mu$, that is by increasing $\mu$ the inverse magnetic catalysis (IMC) turns into magnetic catalysis (MC).
This result is compatible with the observation of IMC in the lattice simulations at vanishing $\mu$ and extends it to finite but small values of $\mu$ which can also be checked on the lattice. 
However, the MC seen at higher values of $\mu$ is beyond the lattice calculations power.
Nevertheless, this result can be served as a confirmation of the observation of \cite{Gursoy} found using a holographic study with a different model and in the chiral transition level.

It is worth noticing that the abovementioned behavior is the common behavior of the solutions of our system regardless of the values of the physical parameters $\mu$ and $B$.
Choosing other values for the free parameters of the system ($c$ and $a$), and/or other forms for the scale factor $A(z)$, however, may lead to different solutions with different phase structures, for example the large/small black hole phase transition/cross over could appear, where a thermodynamically stable black hole solution exists at any given temperature of the system \cite{reconst3,reconst4,reconst5}.
However, for the gravity of our interest one observes that there always exists a Hawking-Page phase transition where a large black hole solution suddenly turns to a thermal AdS solution, for any given values of $\mu$ and $B$, at least in the permitted range. 
The interested reader is referred to \cite{hajilou,hajilou2} to see the discussion of the phase transition in this setup in detail.

In \cite{hajilou} they have also calculated the potential energy of a quark-antiquark pair placed at a fixed distance from each other in the field theory. 
This quantity can be read off from the expectation value of a temporal Wilson loop on a rectangular path with a spatial direction along the separation of the quark-antiquark pair.
It is well known that the potential of such a pair is holographically translated to the extremized on-shell Nambu-Goto (NG) action of an open string in the bulk, hanging from the boundary.
When the system is in the confined phase, quark and antiquark are connected with an open string in the U-shape configuration, in the gravity side.
In the gauge theory side the distance between the quark and antiquark can be increased at will while they are bound to each other and this can be interpreted by a linearly increasing potential.
The linear behavior of the quark potential at large distances indicates the confinement of the theory.
Therefore, the heavy quark potential is one of the most important observables relevant to the confinement.
For bottom-up holographic theories this is realized by the presence of a wall in the bulk.
The tip of an open string hung from the boundary cannot go beyond the wall and as the distance between the quarks is increased, the coresponding string just lay on the wall.
Disappearance of the dynamical wall by changing the parameters of the theory is interpreted as transition to the deconfined phase. 

On the other hand, it is commonly believed that the Hawking-Page phase transition in the gravity side can be translated as the confinement/deconfinement phase transition of the QCD.
In some holographic theories these two kinds of phase transitions may not coincide.
In the system of our interest, by enhancing the value of the magnetic field to values higher than $B_c\simeq 0.37\ GeV$ the dynamical wall disappears, while the background always exhibits the Hawking-Page phase transition at least till we are working in the accepted physical range of the magnetic field.
In such cases it is crucial to realize which one of these transitions corresponds to the confinement/deconfinement phase transition in the gauge theory side.
In the following sections we investigate the critical behavior of the theory when subjected to an external electric field.
Beside being an important problem per se, this critical behavior can also be considered as a key to identify the confinement of the system.

\section{Holographic Schwinger effect}
In order to investigate the effect of a fixed external electric field $E$ on the system of our interest, we calculate the total potential of a quark-antiquark pair under the influence of $E$.
The anisotropy induced by the presence of the magnetic field enables us to analyze the Schwinger effect in anisotropic cases.
The study of the anisotropy influence is important in realistic situations such as the anisotropic media produced by heavy ion collisions in which there are two types of the anisotropy caused by the collision itself and by the magnetic field appeared during the collision.
In our present setup we are able to investigate the effect of the anisotropy caused by the second source.
From this point of view, in what follows, we study the longitudinal (transverse) case where the electric field and the q-$\bar{\mathrm{q}}$ pair are oriented parallel (perpendicular) to the magnetic field direction.

The total potential can be obtained by adding the energy of the interaction with the electric field to the sum of the potential and static energy of the quark-antiquark pair.
As argued in the previous section, the energy of a q-$\bar{\mathrm{q}}$ pair of infinite masses is holographically obtained through the on-shell NG action of an open string with a world-sheet bounded on the AdS boundary by a temporal rectangular Wilson loop with dimensions $l$ and ${\cal T}$, where the time ${\cal T}$ is supposed to be much larger than the fixed distance $l$ between the q and $\bar{\mathrm{q}}$. 
Following the prescription of \cite{semenoff}, to avoid the suppression of the pair production due to considering the quarks of infinite masses, we put a probe D3-brane in an intermediate position $z_0$ of the bulk and the string endpoints are attached to this brane instead of the AdS boundary.

Parametrizing the string coordinates as $t=\tau$, $x_{\parallel}=\sigma$ ($x_{\perp 1}=\sigma$) for the longitudinal (transverse) case, $z=z(\sigma)$, and keeping the other spatial coordinates fixed, the induced metric on the string world-sheet is obtained as
\begin{align}\label{indmetric}
  ds_{s,\mathrm{ind}}^2=\frac{L^2 e^{2{\cal A}_s(z)}}{z^2}\left[-g(z)d\tau^2+\left(f_{\parallel,\perp}(z)+\frac{z'^2}{g(z)}\right)d\sigma^2\right],
\end{align}
in which $f_{\parallel}=1$ and $f_{\perp}=e^{B^2z^2}$ for $\sigma=x_{\parallel}$ and $\sigma=x_{\perp}$, respectively.
Then, by substituting the determinant of this metric into the NG action and obtaining the conserved quantity, the separation length between the quark and antiquark is given by
\begin{align}\label{distance}
l_{\parallel,\perp}=2 \int_{z_0}^{z_c} dz \frac{\sqrt{f_{\parallel,\perp}(z_c)g(z_c)}e^{2{\cal A}_s(z_c)}z^2}{f_{\parallel,\perp}(z)g(z)e^{2{\cal A}_s(z)}z_c^2\sqrt{h_{\parallel,\perp}(z)}},
\end{align}
where $z_c=z(0)$ is the position of the tip of the open string in the bulk, at which $z'=0$, and
\begin{align}\label{hfunc}
h_{\parallel,\perp}(z)=1-\frac{f_{\parallel,\perp}(z_c)g(z_c)e^{4{\cal A}_s(z_c)}z^4}{f_{\parallel,\perp}(z)g(z)e^{4{\cal A}_s(z)}z_c^4}.
\end{align}
Moreover, the total potential is derived as follows:
\begin{align}\label{vtotal}
V_{\mathrm{total}}^{\parallel,\perp}(l_{\parallel,\perp})=2 T_F L^2 \int_{z_0}^{z_c} dz \frac{e^{2{\cal A}_s(z)}}{z^2 \sqrt{h_{\parallel,\perp}(z)}}-E\ l_{\parallel,\perp},
\end{align}
where $T_F$ is the string tension.
\subsection{Critical electric fields}
Previous studies of the holographic Schwinger effect have demonstrated that there exist two critical electric fields labeled as $E_s$ and $E_c$, for a general field theory with gravity dual.
Below $E_s$ no pair production happens, even for the massless particles, since the electric force cannot compensate the confining force between particles.
As the electric field reaches this value, the charged particles begin producing from the vacuum.
This critical behavior is restricted to confining phase and $E_s=0$ for a deconfined theory, i.e., at least the zero-mass particle-antiparticle pairs can be produced by applying any nonzero external electric field.
Between $E_s$ and $E_c$ the Schwinger effect occurs through a quantum tunneling process.
Above $E_c$ which depends on the mass of the particles, the pairs can be created catastrophically and the vacuum would decay.

The value of $E_c$ can be obtained from the DBI action of a D3-brane located at $z_0$ in the bulk, which is written as 
\begin{align}\label{dbiD3}
S_{\mathrm{D3}}=-T_{\mathrm{D3}}\int d^4 x \sqrt{-\det \left(g_{\mu \nu} +{\cal F}_{\mu \nu}\right)},
\end{align}
where $T_{\mathrm{D3}}$ is the D3-brane tension.
In our system this action for the longitudinal and transverse cases is respectively given by
\begin{align}\label{dbiD3L}
S_{\mathrm{D3}}^{\parallel}=-T_{\mathrm{D3}}\int d^4 x \frac{L^2e^{2{\cal A}_s(z_0)+B^2z_0^2}}{z_0^2}\sqrt{\left(\frac{e^{2{\cal A}_s(z_0)}L^2}{z_0^2}\right)^2g(z_0)- \frac{E^2}{T_F^2}},
\end{align}
and 
\begin{align}\label{dbiD3T}
S_{\mathrm{D3}}^{\perp}=-T_{\mathrm{D3}}\int d^4 x \frac{L^2e^{2{\cal A}_s(z_0)}}{z_0^2}\sqrt{\left(\frac{e^{2{\cal A}_s(z_0)}L^2}{z_0^2}\right)^2 g(z_0)e^{2B^2z_0^2} - \frac{E^2}{T_F^2}e^{B^2z_0^2}}.
\end{align}
The action is real provided $E\leqslant E_c$, where
\begin{align}\label{ec}
E_c^{\parallel,\perp}=T_F \frac{e^{2{\cal A}_s(z_0)}L^2}{z_0^2}\sqrt{g(z_0)f_{\parallel,\perp}(z_0)}.
\end{align}
This critical field can also be found using the total potential relation (\ref{vtotal}). 
As one knows, $E_c$ is the critical value of the electric field, at which the total potential barrier vanishes, i.e., at $E_c$ the slope of the total potential tends to zero when the separation length of q and $\bar{\mathrm{q}}$ approaches zero.
To obtain $E_c$ by virtue of this condition, it is helpful to rewrite Eqs.\,(\ref{distance}) and (\ref{vtotal}) in terms of dimensionless parameters $y\equiv \frac{z}{z_c}$ and $b\equiv \frac{z_c}{z_0}$.
Then, similar to \cite{davoud} we calculate the derivative of the total potential with respect to the separation length $l$ by the use of the chain rule and simply find its limit at $l \to 0$ or equivalently $b\to 1$.
The resulting $E_c$ from this calculation is the same as Eq.\,(\ref{ec}).

The other critical electric field, $E_s$, which is only present in the confined phase, can be obtained in a similar way using the potential analysis.
The potential barrier goes to infinity for $E\leqslant E_s$.
In other words, at $E=E_s$ the total potential goes to a constant value as the separation length $l$ goes to infinity or equivalently $z_c \to z_w$, where $z_w$ is the position of the IR cutoff in the bulk.
Notice that in our case a closed form for $E_s$ cannot be found using this condition, although it can be found numerically.
However, we can find $E_s$ through comparison of our metric with the general one in \cite{Sch1} where the formulae of the critical electric fields $E_c$ and $E_s$ have been obtained as universal properties of general confining backgrounds.
This leads to Eq.\,(\ref{ec}) for $E_c$ and the formula of $E_s$ is derived as
\begin{align}\label{es}
E_s^{\parallel,\perp}=T_F \frac{e^{2{\cal A}_s(z_w)}L^2}{z_w^2}\sqrt{g(z_w)f_{\parallel,\perp}(z_w)},
\end{align}
which is nonzero only when $z_w$ has a finite value. 
This relation goes to zero as $z_w$ goes to infinity, characterizing the deconfined phase.
For a holographic confined theory with the IR cutoff $z_w$, the radial position is restricted to the interval $[0,z_w)$ (or $[z_0,z_w)$ in the case of finite-mass quarks).
As explained in \cite{davoud} in detail, the value of $z_w$ can be obtained by using the positivity condition of the function $h(z)$ presented in Eq.\,(\ref{hfunc}), which ensures that the distance in Eq.\,(\ref{distance}) remains real in the whole integral range.
In this way we find the upper bound $z_w$ for $z_c$, i.e., the maximum value that the tip of an open string attached to the D3-brane can reach in the bulk is $z_w$. 
As illustrated there, in practice $z_w$ is found simply by extremizing the function $h(z)$.

Here we explain another interesting and easy way to find the IR wall position.
As we know the Schwinger effect does not occur when the electric field is weaker than $E_s$, i.e., the minimum value of the effective string tension which is reached at the IR cut-off.
Hence, for our purpose, we can use the relation of $E_s$ as a function of $z$ as 
\begin{align}\label{ecrit}
E_{\mathrm{critical}}^{\parallel,\perp}(z)=T_F \frac{e^{2{\cal A}_s(z)}L^2}{z^2}\sqrt{g(z)f_{\parallel,\perp}(z)}.
\end{align}
It is easy to see that $E_{\mathrm{critical}}(z_0)=E_c$ and $E_{\mathrm{critical}}(z_w)=E_s$.
Since $E_s$ is the absolute minimum value of the above function or the minimum value of the effective string tension, one concludes that $z_w$ can be easily found via minimizing the function $E_{\mathrm{critical}}(z)$ with respect to $z$.


Now, we investigate the critical behavior of our system under the effect of the electric field $E$ by depicting some graphs, through which we can also study the possible phases of our field theory.
Notice that in all the graphs throughout the paper the critical electric fields and the total potential are rescaled with $T_F L^2$.
As mentioned before, the study of the thermodynamics shows two phases for our employed geometry in the gravity side: the thermal AdS phase and the black hole phase.
We now pursue the confinement and deconfinement phases of the field theory side considering the relations of the critical electric fields in any of the abovementioned phases.

\begin{figure}[h]
\begin{center}
\includegraphics[width=6.8cm]{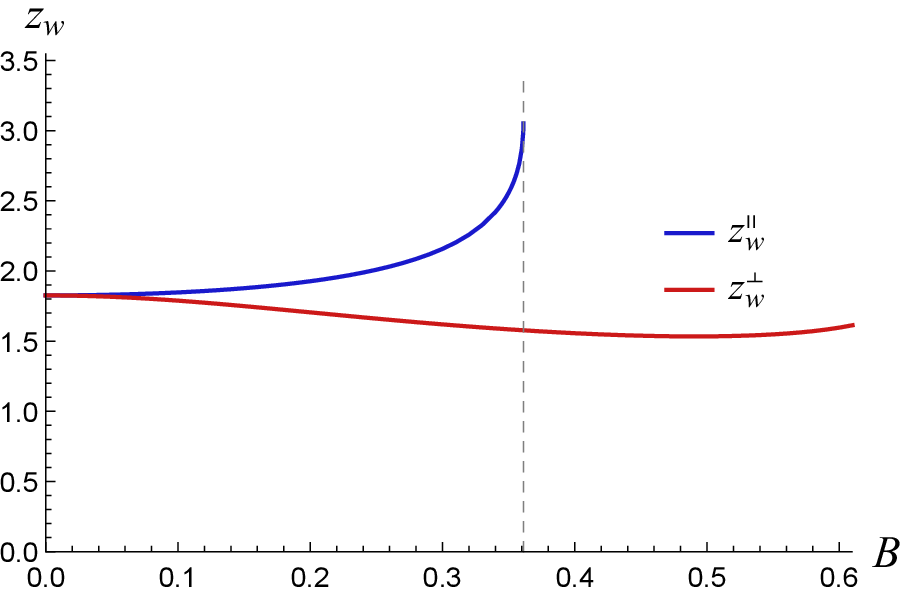}\hspace{0.3cm}
\includegraphics[width=6.8cm]{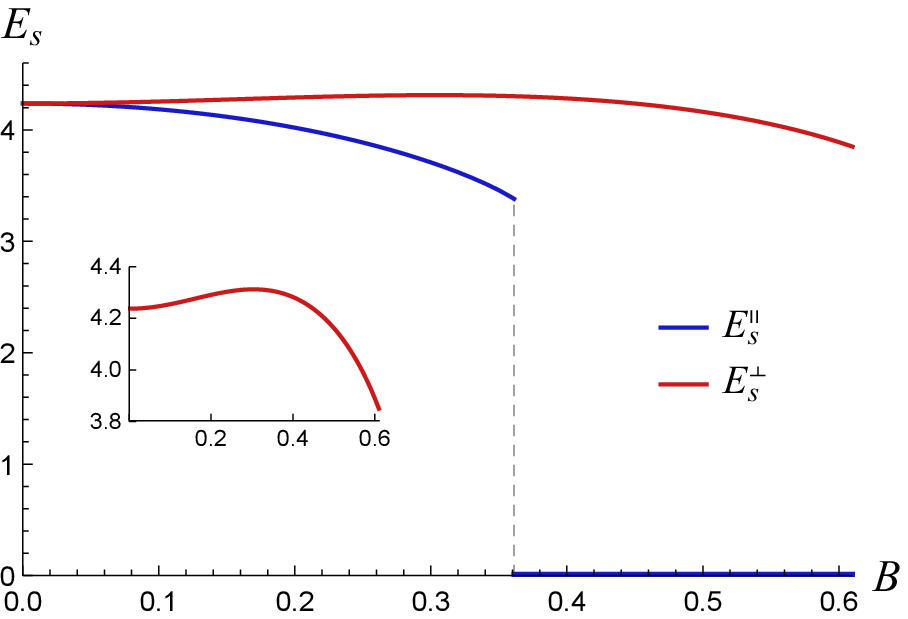}
\end{center}
\caption{\footnotesize 
Left (right) graph shows the position of the IR wall (the critical electric field $E_s$) as a function of the magnitude of the magnetic field for both cases with the electric field parallel and perpendicular to the magnetic field direction.}
\label{esthermal}
\end{figure} 

\begin{figure}[h]
\begin{center}
\includegraphics[width=6.8cm]{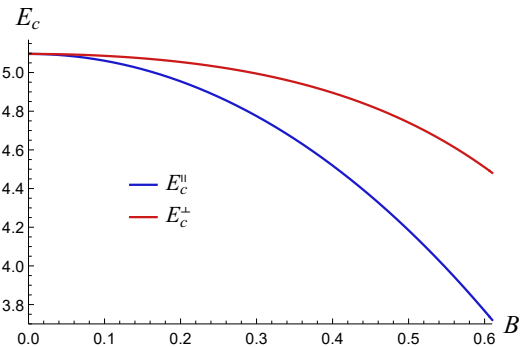}
\end{center}
\caption{\footnotesize 
Critical electric field $E_c$ as a function of the magnetic field for the cases with parallel and perpendicular electric fields in the thermal AdS phase.}
\label{ecthermal}
\end{figure}

In the thermal AdS phase where no black hole is present, Eqs.\,(\ref{ec}) and (\ref{es}) are both independent of the parameters $\mu$ and $z_h$. 
In this case $g(z)=1$ meaning $z_h \to \infty$. 
Figure \ref{esthermal} depicts the critical electric field $E_s$ as a function of $B$, for both longitudinal and transverse cases where the electric field and q-$\bar{\mathrm{q}}$ pair are parallel and perpendicular to the magnetic field direction, respectively.
The position of the IR cutoff $z_w$ is also presented in the left graph of this figure.
As can be seen, while in the transverse case (red curves), for each physically accepted $B$, there exists an IR cutoff and consequently a finite $E_s$, in the longitudinal case (blue curves) $E_s$ suddenly jumps to zero at $B_{\mathrm{critical}}\simeq 0.36118\ GeV$.
This means that the quarks do not feel the confining force in the direction of the magnetic field with the magnitude greater than the mentioned value.
$E_s^{\parallel}$ decreases with $B$, however, $E_s^{\perp}$ increases for small values of $B$ and then start decreasing by further increase of $B$.
The behavior of $E_c$ as a function of $B$ is depicted in Fig.\,\ref{ecthermal} for the longitudinal and transverse cases, showing the decrease of $E_c$ with the increase of $B$.

\begin{figure}[h]
\begin{center}
\includegraphics[width=6.8cm]{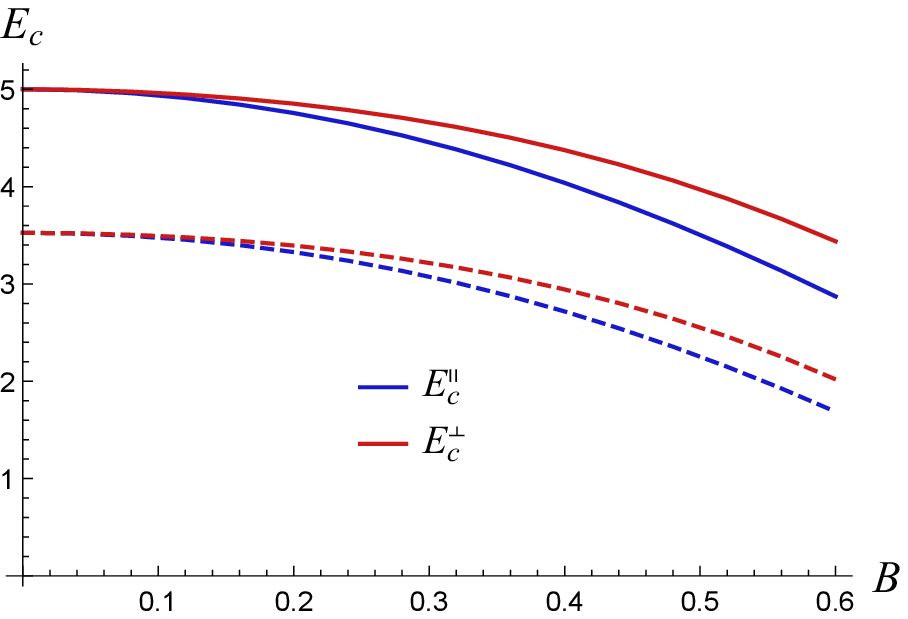}\hspace{0.3cm}
\includegraphics[width=6.8cm]{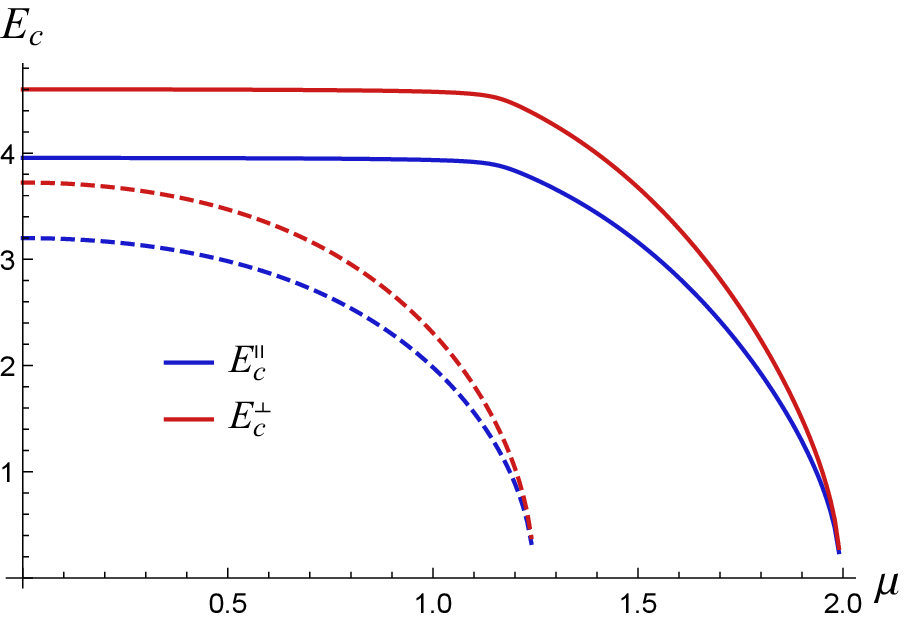}
\end{center}
\caption{\footnotesize 
Left graph: $E_c$ versus $B$ in the longitudinal and transverse cases, for $\mu=0$ (solid lines) and $\mu=1$ (dashed lines).
Right graph: $E_c$ versus $\mu$ in the longitudinal and transverse cases at $B=0.55$.
All the graphs are in the black hole phase of the gravity corresponding to the deconfined phase of the field theory with the temperature $T=0.27$ for the left graph, and $T=0.17$ and $T=0.27$ for the solid and dashed lines, respectively, in the right graph.}
\label{ecbh}
\end{figure}

We now turn to consider the critical electric fields in the black hole phase.
As expected, in this case there is no IR wall in the bulk, reflecting the deconfinement in the field theory side and consequently $E_s=0$.
The critical electric field $E_c$ is displayed in the left panel of Fig.\,\ref{ecbh} for $\mu=0,1$ (solid and dashed lines, respectively) for both longitudinal and transverse cases (blue and red lines, respectively) at $B=0.55\ GeV$ and $T=0.27\ GeV$.
The right panel of this figure shows $E_c$ versus $\mu$ at $B=0.55\ GeV$ and the temperature is chosen to be $T=0.17\ GeV$  and $T=0.27\ GeV$ for the solid and dashed lines, respectively.
As can be seen, both $B$ and $\mu$ reduce the critical field $E_c$.
Another important result here is that while the increase of $B$ induces anisotropy, $\mu$ works in the opposite direction and reduces the effect of the anisotropy induced by the presence of the magnetic field. 
From the right panel we furthermore infer that the increase of the temperature reduces the critical field $E_c$ and also favors the Schwinger effect, as expected.
The higher the temperature, the less the chemical potential at which $E_c$ vanishes and the vacuum starts decaying.

\begin{figure}[h]
\begin{center}
\includegraphics[width=6.8cm]{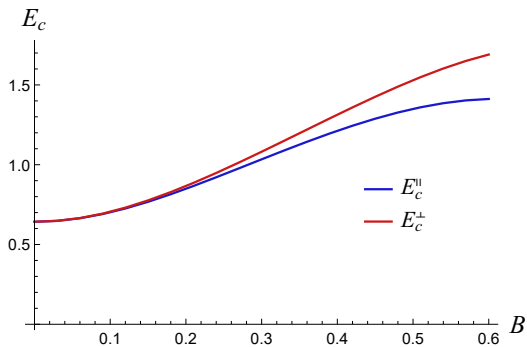}
\end{center}
\caption{\footnotesize 
$E_c$ versus $B$ in the longitudinal and transverse cases, for $\mu=2$.
These graphs are in the black hole phase of the gravity corresponding to the deconfined phase of the field theory with the temperature $T=0.15$.}
\label{ecbh2}
\end{figure}

We should stress an interesting point here.
Although $E_c$ decreases by increasing the magnetic field for low values of the chemical potential, the situation changes as $\mu$ increases enough (Fig.\,\ref{ecbh2}).
This result reminds us of the alternation of the inverse magnetic catalysis to magnetic catalysis by going to the higher values of $\mu$ as seen in Fig.\,\ref{T-HPB}.

\subsection{Potential analysis}

We now turn our attention to the analysis of the total potential 
of q-$\bar{\mathrm{q}}$ under the influence of the external electric field, evaluated in Eq.\,(\ref{vtotal}).
We consider the Schwinger effect in different phases of the background, separately.
\subsubsection*{Thermal AdS phase}

Let us first focus on the separation length of the quark-antiquark pair, given in Eq.\,(\ref{distance}) and analyze it in different situations.

Before proceeding to explain the results, notice that in some of the following graphs we use the parameter $a$.
Here $a=\frac{z_0}{z_c}$ is the inverse of the position that the tip of the open string can reach in the bulk, rescaled with respect to $z_0$. 
The farthest point attainable by the turning point of the open string is $z_c=z_{\mathrm{IR}}$.  
In the thermal AdS phase for the cases where there exists an IR wall cutting off the radial position at a finite value $z_w$, that is for a pair of quarks in a direction transverse to the magnetic field (with physically acceptable value) and a pair of quarks in the same direction as the magnetic field with the value $B\leqslant B_{\mathrm{critical}}$, we have $z_{\mathrm{IR}}=z_w$.
However, for the longitudinal case with $B> B_{\mathrm{critical}}$, where suddenly the IR wall disappears, $z_{\mathrm{IR}}\to \infty$.
In the black hole phase of the background, there always exists a horizon at a finite radial position $z_h$ in both longitudinal and transverse cases and for the whole range of the parameters $\mu$ and $B$, and consequently $z_{\mathrm{IR}}=z_h$.

\begin{figure}[h]
\begin{center}
\includegraphics[width=6.8cm]{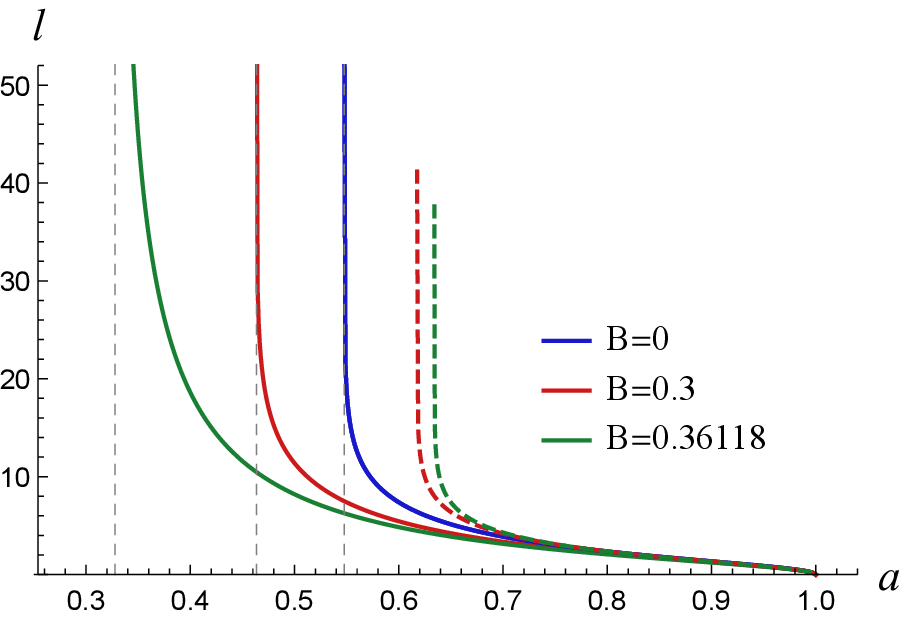}\hspace{0.3cm}
\includegraphics[width=6.8cm]{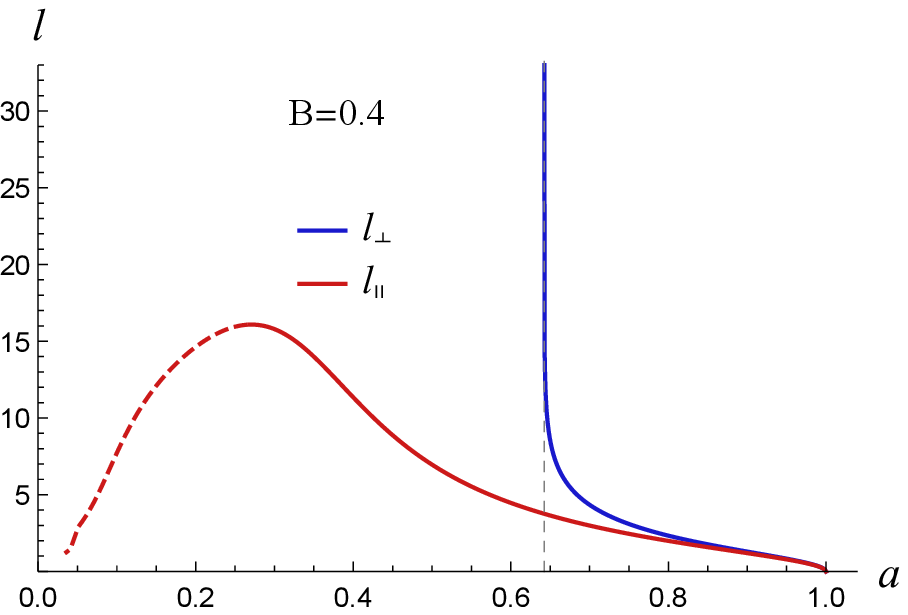}
\end{center}
\caption{\footnotesize 
The separation length of q-$\bar{\mathrm{q}}$ pair versus $a$ for various values of the magnetic field in the thermal AdS phase.}
\label{laconf}
\end{figure} 

Figure \ref{laconf} shows the distance between the quark and antiquark as a function of the parameter $a$.
In the left panel we display the separation length for various values of the magnetic field, obeying the condition $B\leqslant B_{\mathrm{critical}}$.
At $B=0$, the absence of a preferred direction results in the equality of $l_{\parallel}$ and $l_{\perp}$, depicted by the solid blue line.
Moreover, the other solid (dashed) lines refer to the sample longitudinal (transverse) cases.
As can be simply observed, in all cases the tip of the string cannot exceed a minimum value of $a$  ($a_{\mathrm{min}}=\frac{z_0}{z_w}$) at which the distance between the quarks asymptotes the infinity and the string lies on the IR wall, which is a characteristic of confining theories.
In the right panel, the blue (red) line shows $l_{\perp}$ ($l_{\parallel}$) for a magnetic field greater than $B_{\mathrm{critical}}$.
While the behavior of $l_{\perp}$ indicates the confinement, the existence of a maximum value for $l_{\parallel}$ confirms the deconfinement of the field theory in the longitudinal direction, as expected according to the discussion of the previous section.
In general, there are three possible string configurations: two U--shape connected configurations and one disconnected solution where two open strings stretch from the UV boundary (in our case the D3-brane at $z_0$) to $z_{\mathrm{IR}}$ which goes to infinity in the present case.
At $B> B_{\mathrm{critical}}$, for $l_{\parallel} \leqslant l_{\parallel}^{\mathrm{max}}$ the stable solution (with lower energy) is the U--shape configuration for which the turning point of the string is closer to the boundary (lower $z_c$ or higher $a$), represented by the solid red line in the right panel of 
Fig.\,\ref{laconf}. 
For $l_{\parallel} > l_{\parallel}^{\mathrm{max}}$ the favored solution is the disconnected one indicating the detachment of the q and $\bar{\mathrm{q}}$ in the field theory side. 

\begin{figure}[h]
\begin{center}
\includegraphics[width=6.8cm]{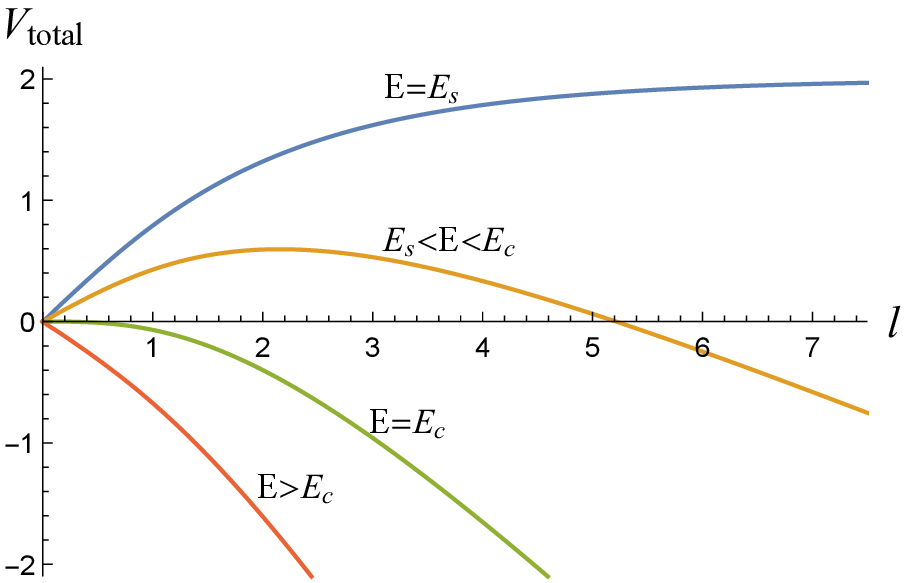}\hspace{0.3cm}
\includegraphics[width=6.8cm]{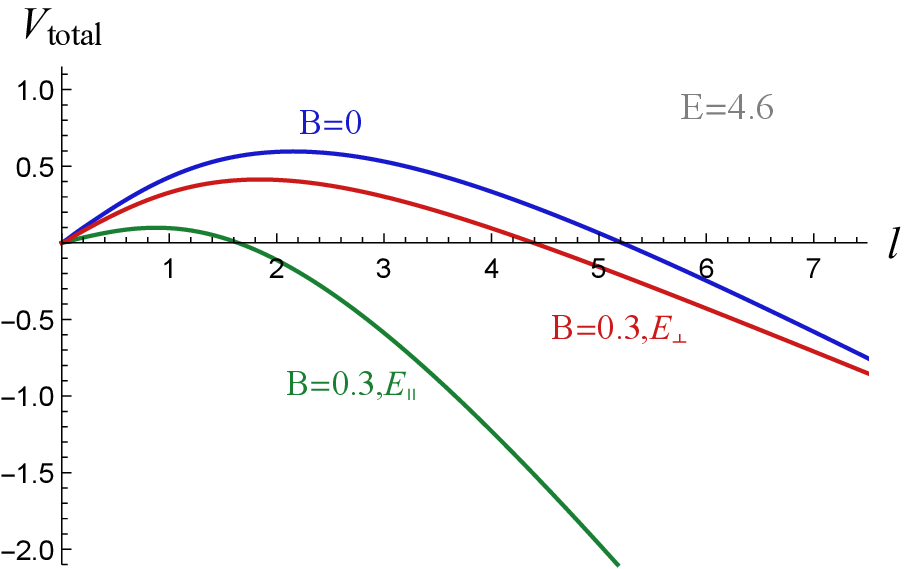}
\end{center}
\caption{\footnotesize 
The Total potential versus the separation length in the thermal AdS phase. Left panel shows the potentials for various values of the electric field at $B=0$ and right panel considers the effect of the magnetic field on the potential at $E=4.6$ in both longitudinal and transverse cases.}
\label{VLconf}
\end{figure} 

Let us now focus on the analysis of the total potential in the thermal AdS phase of the background. 
As the plot in the left panel of Fig.\,\ref{VLconf} shows for the total potential at $B=0$ versus $l$, the potential barrier varies as the electric field varies. 
At $E=E_s$ the potential barrier becomes constant as the distance between the quarks tends to infinity. 
For higher values of the electric field the barrier decreases and becomes finite so that the quarks can be liberated through a tunneling process. 
At $E=E_c$ the barrier vanishes completely for quarks of mass $m$. 
Therefore, for $E>E_c$ the vacuum decays catastrophically. 
This figure confirms the previous results reported for the Schwinger effect studied by the potential analysis.
The right graph of this figure compares the total potential without magnetic field, and with magnetic field for longitudinal and transverse cases at a sample electric field $E=4.6\ GeV^2$. 
The graphs of both longitudinal and transverse cases show a decrease in the height and width of the potential barrier.
Since the Schwinger effect happens through a tunneling process, this behavior means that the quarks can be freed more easily in the presence of the magnetic field.
However, 
Notice that in the longitudinal case the quarks are faced with a smaller barrier. Therefore, the application of the magnetic field in the same direction as the electric field has more impact on the enhancement of the Schwinger effect.
\subsubsection*{Black hole phase}
\begin{figure}[h]
\begin{center}
\includegraphics[width=6.8cm]{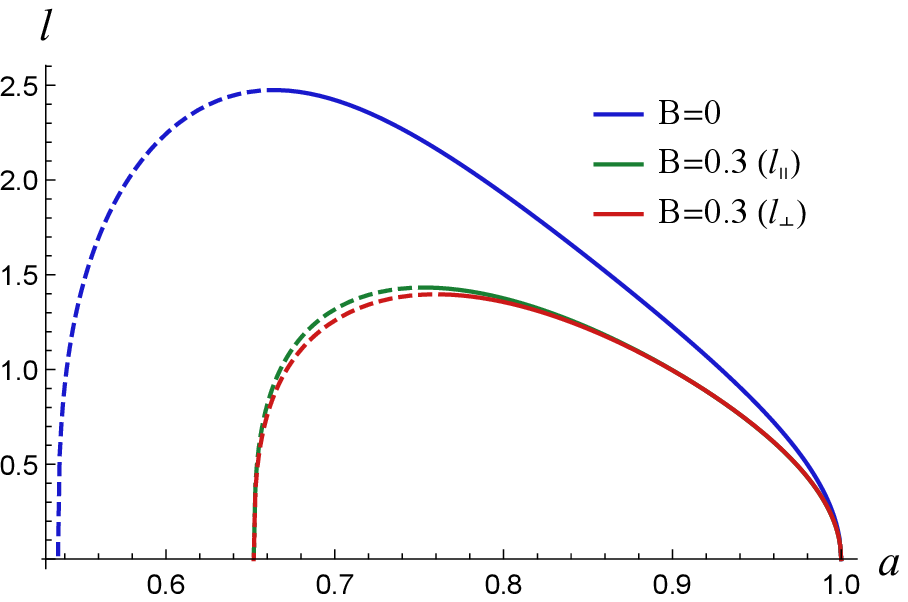}\hspace{0.3cm}
\includegraphics[width=6.8cm]{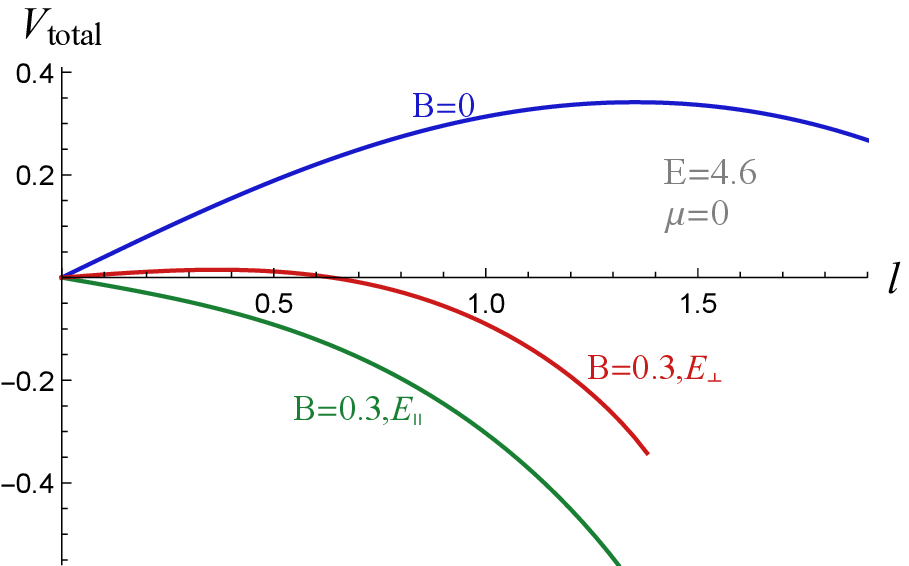}
\end{center}
\caption{\footnotesize 
Left and right panel, respectively, depict the separation length of q-$\bar{\mathrm{q}}$ pair versus $a$ and the total potential versus $l$ with $\mu=0$, in the presence and absence of the magnetic field, in the black hole phase at $T=0.27$.}
\label{lvBH1}
\end{figure}

\begin{figure}[h]
\begin{center}
\includegraphics[width=6.8cm]{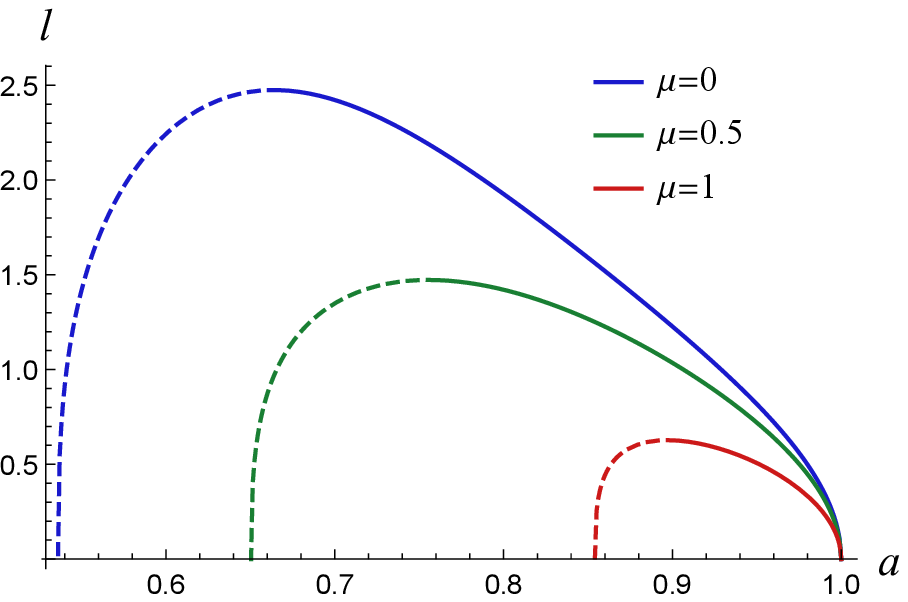}\hspace{0.3cm}
\includegraphics[width=6.8cm]{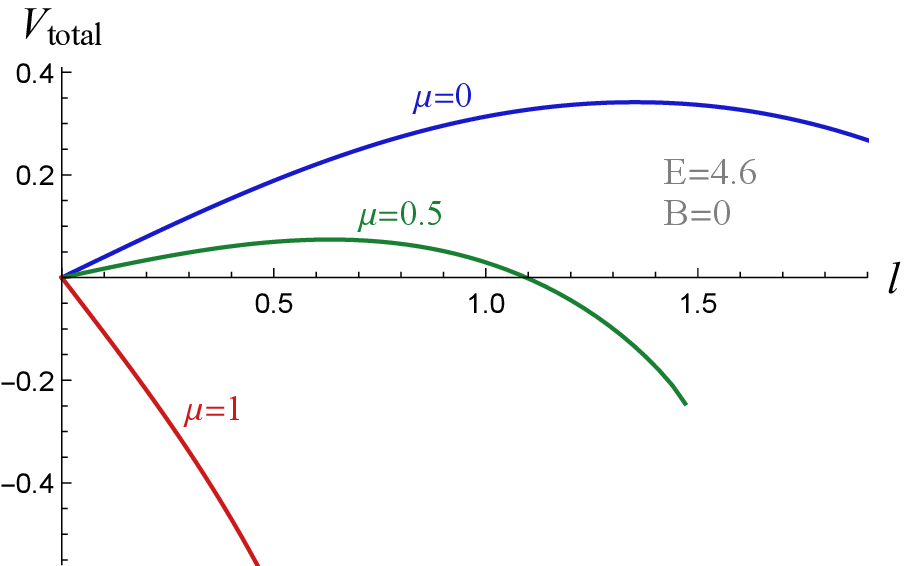}
\end{center}
\caption{\footnotesize 
Left and right panel, respectively, show the separation length of q-$\bar{\mathrm{q}}$ pair versus $a$ and the total potential versus $l$ for various values of the chemical potential in the black hole phase at $T=0.27$ and $B=0$.}
\label{lvBH2}
\end{figure}

\begin{figure}[h]
\begin{center}
\includegraphics[width=6.8cm]{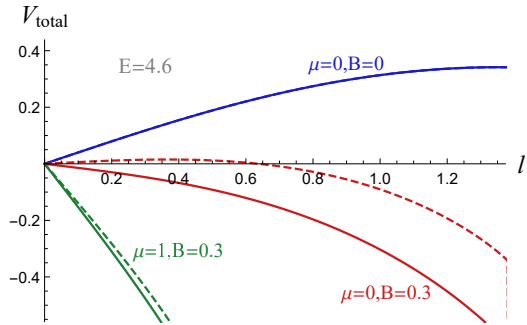}
\end{center}
\caption{\footnotesize 
The total potential versus $l$ for various values of the chemical potential in the black hole phase at $T=0.27$. Solid and dashed lines refer to longitudinal and transverse directions of the electric field, respectively.}
\label{lvBH3}
\end{figure}

\begin{figure}[h]
\begin{center}
\includegraphics[width=6.8cm]{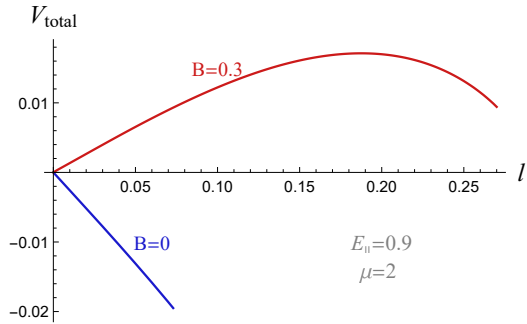}
\end{center}
\caption{\footnotesize 
The total potential versus $l$ in the black hole phase at $T=0.15$ for a system with $\mu=2$. The electric field is chosen to be $E=0.9$.}
\label{lvBH4}
\end{figure}

Now we report the results for the black hole phase which are illustrated briefly through Figs.\,\ref{lvBH1}--\ref{lvBH4}.
The left and right panels of Fig.\,\ref{lvBH1} depict respectively the separation length of the quark and antiquark versus the parameter $a$ and the total potential versus $l$, at  $\mu=0$, $T=0.27\ GeV$, and for various values of the magnetic field with different orientations with respect to the direction of the separation of the quarks.
As it is apparent from the left graph, the pair of the quarks becomes free at a finite value of $a$ or equivalently a finite  horizon position $z_h$.
Such a behavior confirms the deconfining character of the field theory, as expected for the black hole phase according to our previous discussion.
We can also see from this graph that the presence of the magnetic field reduces the $l_{\mathrm{max}}$ at which the quarks become liberated, regardless of its orientation. 
Hence, one expects that the presence of the magnetic field helps the quarks to be freed easier.

In the right panel of Fig.\,\ref{lvBH1} we compare the total potential for three cases, i.e., in the absence of the magnetic field $B=0$, in the presence of the magnetic field with the electric field parallel and perpendicular to the magnetic field direction.
In all these cases $\mu=0$ and $T=0.27\ GeV$, and we have chosen the sample electric field $E=4.6\ GeV^2$.
As expected, the presence of the magnetic field enhances the Schwinger effect by reducing the potential barrier that the quarks are faced with. 
This effect is more evident when the magnetic and electric fields are applied at the same direction.

Figure \ref{lvBH2} considers the effect of changing the chemical potential while the magnetic field is turned off.
According to these graphs, the effect of the chemical potential is qualitatively similar to the magnetic field, i.e., it works in favor of the pair production from the vacuum.
To explore the effect of simultaneous presence of $\mu$ and $B$, we can consider the graph in Fig.\,\ref{lvBH3} where we show the total potential as a function of $l$ for $\mu=0,1$ and $B=0,0.3\ GeV$ at $T=0.27\ GeV$.
The electric field is chosen to be $E=4.6\ GeV^2$.
In the presence of $B$, the cases with the electric field parallel and perpendicular to the direction of $B$ are displayed with solid and dashed lines, respectively.
One can see that both $\mu$ and $B$ simplify the creation of the real quarks from the vacuum, as they both decrease the height and width of the potential barrier.
We moreover see that the chemical potential reduces the anisotropy induced by the magnetic field, consistent with the result found by the study of the critical electric field $E_c$.
From Fig.\,\ref{lvBH4} we can deduce that at high enough values of $\mu$ the increase of $B$ would no longer leads to the decrease of the height and width of the potential barrier.
Instead, at these values of $\mu$ the magnetic field reduces the Schwinger effect, which is consistent with the result seen in Fig.\,\ref{ecbh2} showing the increase of the critical behavior $E_c$ by the magnetic field at high enough $\mu$. 
In the next section we summarize our study and present a discussion about the results.

\section{Summary and conclusion}
Most of the holographic Schwinger effect studies have been done through top-down holographic models or bottom-up theories which are not the solution of the Einstein equation while choosing a self-contained model with the most similarity to the real QCD is very important to obtain results comparable with the experimental situations and/or gain knowledge about the theory itself.
To that purpose, we have employed a bottom-up holographic model which is an analytic solution of the equations of motion, derived and explained in \cite{hajilou} to investigate the effect of a finite external electric field on a confining QCD-like theory. This model called the magnetized EMD model contains the chemical potential, a running dilaton and a background magnetic field responsible for the anisotropy in different spatial directions. The scale factor of the ansatz metric has been chosen to be quadratic with one free parameter which along with other free parameters of the gravity solution is determined by comparison with some experimental and lattice simulation results.
This choice leads to an analytic solution with a Hawking-Page phase transition for any given values of $\mu$ and $B$; there is a black hole solution in the gravity side corresponding to the field theory with a temperature higher than $T_{\mathrm{HP}}$ and a thermal AdS solution corresponding to lower temperatures. 
We have furthermore realized that by increasing $\mu$ to sufficiently large values the inverse magnetic catalysis observed for zero or small $\mu$ turns into magnetic catalysis. 
By investigating the behavior of the theory under the influence of an electric field we could elaborate more on the confinement of the field theory side at each phase of the gravity solution.

Applying the total potential analysis, we have been able to calculate the critical electric field $E_s$, the threshold value for starting the Schwinger effect, and $E_c$ above which the creation of the particles occurs catastrophically and without any obstacle. 
We have examined the behavior of these critical fields along with the separation length of the quark-antiquark pair and the total potential in two phases of the background gravity as the physical parameters change. 
Here we report the main results.

It has been realized that in the thermal AdS background, not for all the values of the magnetic field is there a critical electric field $E_s$. 
Since the existence of a nonzero value for $E_s$ is a characteristic of the confinement, the thermal AdS background does not always correspond to the confining phase, and according to the calculations, the behavior of the dual field theory, instead, mimics the deconfinement for $B>B_{\mathrm{critical}}$, when the electric and magnetic fields are at the same direction (the longitudinal case). 
This conclusion has been also observed in the behavior of the separation length as a function of the position of the string turning point.
In the thermal AdS background the critical electric fields, the separation length and the total potential expression do not depend on $\mu$, meaning that the value of $\mu$ does not affect the pair production. 
However, we have learned that the magnetic field facilitates the Schwinger effect, in most of the cases.
$E_s^{\parallel}$ decreases with $B$. $E_s^{\perp}$ first increases with $B$, then by further increasing $B$ it decreases. 
$E_c$ for both longitudinal and transverse directions decreases with $B$. 
Furthermore, the enhancement of $B$ reduces the height and width of the potential barrier the quarks are facing with, indicating more chance for the production of the quarks.
A similar effect of the magnetic field has been observed for zero or sufficiently small values of the chemical potential, in the black hole phase.
The magnetic field works in favor of the Schwinger effect for both longitudinal and transverse cases, which is resulted from both $E_c$ and the total potential barrier. 
The field theory dual to this phase is always deconfined, and thereupon $E_s=0$.
However, for high enough values of $\mu$ the inverse magnetic catalysis (decrease of $E_c$ with $B$) turns into the magnetic catalysis.
We have moreover found that the chemical potential works in favor of the Schwinger effect and also reduces the distinction between the graphs for the two directions, caused by the anisotropy induced by $B$.
Our calculations have also shown that the effect of $B$ on the production of the quarks aligned parallel to $B$ is more than those perpendicular to $B$. This has been observed both in the critical electric fields and the total potential barrier and in both phases.
The results found here for the effect of $B$ on the Schwinger effect are at least qualitatively consistent with the ones found in \cite{magnetized} for the weak magnetic field solutions. 
Our calculations have shown that one could find a higher chance of producing quarks by applying magnetic fields in most of the cases, meaning that the magnetic field can be used to make this elusive effect more observable.

In \cite{magnetized} they consider a magnetized Einstein-Maxwell model dual to the deconfined phase of a field theory with a background constant magnetic field. Therefore, it takes into account the back reaction of the magnetic field on the geometry as is the case with our theory.
However, our findings have considerable differences with those obtained in works such as \cite{Sch2,Sch3,davoud} and \cite{magneticdecay} where they study the Schwinger effect in the presence of an external magnetic field with no back reaction on the theory, using the analysis of the total potential of a quark pair and/or the calculation of the production rate from the circular Wilson loop, and the study of the stability induced by the electromagnetic field using the imaginary part of the effective DBI-action of a probe D7-brane in the gravity side, respectively. 
The models used in these studies are either top-down models or bottom-up models whose their metric is not a solution of the equations of motion.
In most of the situations, e.g., when the electric and magnetic fields are perpendicular to each other for both approaches and in the parallel case for the potential approach, the magnetic field does not enhance the chance of the pair production, unlike the case we studied.
An interesting direction worth following in future works is to find the source of this contradiction, which can be sought either from the back reaction effect of the field or from the type of model used.

As we mentioned, by choosing a different function for the scale factor of the ansatz metric, the phase structure of the background could be different. 
Therefore, it could have interesting consequences to compare the possible phases when different scale factors are chosen, by which we can search for the most similar holographic confining model to real QCD and also study the response of such a theory to electromagnetic fields.


 \end{document}